\documentclass[preprint, prX]{revtex4}
\usepackage[all]{xy}
\usepackage{amssymb, amsmath}
\usepackage{accents}
\usepackage{graphicx}
\usepackage{subfig}

\begin{document}

\title{Quartic Poisson algebras and quartic associative algebras and realizations as deformed oscillator algebras}
\author{ Ian Marquette}
\affiliation{School of Mathematics and Physics, The University of Queensland, Brisbane, QLD 4072, Australia}
\email{i.marquette@uq.edu.au}

\begin{abstract}
We introduce the most general quartic Poisson algebra generated by a second and a fourth order integral of motion of a 2D superintegrable classical system.
We obtain the corresponding quartic ( associative ) algebra for the quantum analog and we extend Daskaloyannis' construction in obtained in context of quadratic algebras and we obtain the realizations as deformed oscillator algebras for this quartic algebra. We obtain the Casimir operator and discuss how these realizations allow to obtain the finite dimensional unitary irreductible representations of quartic algebras and obtain algebraically the degenerate energy spectrum of superintegrable systems. We apply the construction and the formula obtained for the structure function on a superintegrable system related to type I Laguerre exceptionnal orthogonal polynomials introduced recently.
\end{abstract}
\maketitle

\section{Introduction}

In litterature a quadratic associative algebra with three generators ($A$, $B$ and $C$) of a superintegrable system refers to the following mathematical object \cite{Gra1,Zhe1,Gra2,Gra3,Gra4,Zhe2,Das1,Das2,Vin1,Das3,Das4,Mil1,Kre1,Das5,Das6,Que1,Das7,Pos1,Pos2,Pos3} given by~\eqref{quad1},~\eqref{quad2} and~\eqref{quad3}

\begin{subequations}
\begin{equation}
[A,B]=C,   \label{quad1}
\end{equation}
\begin{equation}
 [A,C]=  \alpha A^{2} + \beta \{A,B\} + \gamma A + \delta B + \epsilon ,\label{quad2}
\end{equation}
\begin{equation} 
 [B,C]=  \nu A^{2} + \xi A -\beta B^{2} -\delta B -\alpha \{A,B\} +\zeta, \label{quad3}
\end{equation}
\end{subequations}

where $[,]$ is the commutator $[A,B]=AB-BA$, $\{,\}$ the anticommutator ${A,B}=AB+BA$ and $A$ and $B$ are integrals of motion of second order in momenta of a given Hamiltonian. The structure constants are constrained by the Jacobi identity and can be polynomials of the Hamiltonian in 2D and other integrals of an Abelian subalgebra in application to higher dimensional systems. This algebraic structure is refered as quadratic Racah algebra $QR(3)$ in the particular case $\nu=0$. They is a classical analog i.e. a quadratic Poisson algebra where the comutator is replaced by the Poisson bracket. Let us mention that a quadratic algebra with four generators was discovered earlier in the eigties by Sklyanin \cite{Skl1,Skl2} in the context of the Yang Baxter equation.  The two Casimir operators were also obtained and the representations studied. 

Later, it was shown how the representations of quadratic algebras can be obtained in a systematic manner \cite{Das4} using deformed oscillator algebras \cite{Das8,Que3}. In this approach the finite dimensional unitary irreductible representations of the quadratic algebra are obtained using the structure function of a deformed oscillator algebra and appropriate constraints.

More recently cubic Poisson algebras \cite{Mar1} and cubic associative algebras \cite{Mar2,Mar3} were introduced in context of the superintegrable systems with a second and a third order integrals of motion. They were applied to a class of such systems separable in Cartesian coordinates and used to obtain algebraically the energy spectrum and explain the degeneracies. For a such algebraic structure the modification to the quadratic algebra is minimal and only a cubic term (i.e. $A^{3}$) was added to the right side of equation~\eqref{quad3} and it was observed that a such algebraic structure also allow realizations as deformed oscillator algebras. 

Recently new approaches to construct polynomial algebras for superintegrable systems were introduced using ladder operators \cite{Mar4,Mar5,Kal1,Kal2,Pos4,Que4,Que5} for specific families of systems. In the case of systems with separation of variables in Cartesian coordinates it was observed that the resulting polynomial algebra can be realized as deformed oscillator algebras. However, the existence and systematic study of realizations as deformed oscillator algebras for higher order polynomial algebra is an unexplored subject.

Quadratic, cubic and more generally polynomial algebras are rich and interesting object that have applications in context of superintegrable systems, but as other algebraic structure as Lie algebras they could find applications in other context. The purpose of this paper is to investigate the case of quartic Poisson and in particular quartic algebras and the realizations in term of deformed oscillator algebras. Using this approach we intend to provide explicit formula that would allow to construct finite dimensional unitary representations for quartic algebras. 

Let us present the organisation of the paper. In Section 2, we present the most general quartic Poisson algebra and calculate the Casimir operator. In Section 3, we obtain the quantum analog, a quartic associative algebra. We calculate the Casimir operator and obtained the realizations as deformed oscillator algebras. These realizations allow to construct the Fock type representation of the quartic algebra. In Section 5, in order to illustrate the application of these results in context of superintegrable systems, we apply this construction on a superintegrable system studied recently and related to Laguerre EOP \cite{Que4}. We obtain algebraically the energy spectrum and the degeneracies.

\section{Quartic Poisson algebras}

We consider a two-dimensional Superintegrable Hamiltonian $H$ with a second and fourth order well defined integrals of motion (respectively A and B). 

\begin{equation}
A=\sum_{i+j \leq 2}f_{ik}(x_{1},x_{2})p_{1}^{i}p_{2}^{j}, \label{intA}
\end{equation}
\begin{equation}
B=\sum_{i+j \leq 4}g_{ik}(x_{1},x_{2})p_{1}^{i}p_{2}^{j}, \label{intB}
\end{equation}

where $f_{ik}(\vec{x})$ and $g_{ik}(\vec{x})$ are some unknown functions. We thus have the following Poisson Bracket

\begin{equation}
\{H,A\}_{p}=\{H,B\}_{p}=0. \label{commu}
\end{equation}

The most general quartic Poisson algebra generated by such second order and fourth order integrals has the form :

\begin{subequations}
\begin{equation}
\{A,B\}_{p}=C, \label{quartcl1}
\end{equation}
\begin{equation}
 \{A,C\}_{p}= \tau A^{3} + \alpha A^{2} + 2 \beta AB + \gamma A + \delta B + \epsilon, \label{quartcl2}
\end{equation}
\begin{equation}
 \{B,C\}_{p}= \lambda A^{4} + \mu A^{3} + \nu A^{2} + \xi A + \rho B^{2} +\eta B + 2 \omega A^{2}B + 2 \sigma AB +\zeta. \label{quartcl3}
\end{equation}
\end{subequations}

The right side of the equation~\eqref{quartcl2} and~\eqref{quartcl3} are obtained by considering the power on the left side in terms of the momentum. Thus we form the most general polynomial in $A$ and $B$ to this given order. Let us mention however this is not guaranteed that in general a second and fouth order integrals closed in a such algebraic structure. As observed for many example of superintegrable systems, sometime we need to construct higher order integrals and the corresponding higher order polynomial. 

From the Jacobi equation we have the constraint

\begin{equation}
\{A,\{B,C\}_{p}\}_{p}=\{B,\{A,C\}_{p}\}_{p}. \label{jacobic}
\end{equation}

and we have the relations:

\begin{equation}
\omega=-\frac{3}{2}\tau,\quad \sigma=-\alpha,\quad \rho=-\beta,\quad \eta=-\gamma. \label{jacobiccon}
\end{equation}

The quartic Poisson algebra takes thus the form : 

\begin{subequations}
\begin{equation}
\{A,B\}_{p}=C, \label{quartcl1v2}
\end{equation}
\begin{equation}
 \{A,C\}_{p}= \tau A^{3} + \alpha A^{2} + 2 \beta AB + \gamma A + \delta B + \epsilon, \label{quartcl2v2}
\end{equation}
\begin{equation}
 \{B,C\}_{p}= \lambda A^{4} + \mu A^{3} + \nu A^{2} + \xi A -\beta B^{2} -\gamma B -3\tau A^{2}B -2\alpha AB +\zeta, \label{quartcl3v2}
\end{equation}
\end{subequations}

with $\tau$, $\lambda$ and $\omega$ are constants and the other parameters are given by
\begin{equation}
\alpha=\alpha(H)=\alpha_{0}+\alpha_{1}H,\quad \gamma=\gamma(H)=\gamma_{0}+\gamma_{1}H+\gamma_{2}H^{2}, \quad \delta=\delta(H)=\delta_{0}+\delta_{1}H  \label{structureclas}
\end{equation}
\[\epsilon=\epsilon(H)=\epsilon_{0}+\epsilon_{1}H+\epsilon_{2}H^{2}+\epsilon_{3}H^{3},\quad \mu=\mu(H)=\mu_{0}+\mu_{1}H,\quad \nu=\nu(H)=\nu_{0}+\nu_{1}H+\nu_{2}H^{2} \]
\[ \xi=\xi(H)=\xi_{0}+\xi_{1}H+\xi_{2}H^{2}+\xi_{3}H^{3},\quad \rho=\rho(H)=\rho_{0}+\rho_{1}H,\quad \eta=\eta(H)=\eta_{0}+\eta_{1}H+\eta_{2}H^{2}\]
\[ \sigma=\sigma(H)=\sigma_{0}+\sigma_{1}H,\quad \zeta=\zeta(H)=\zeta_{0}+\zeta_{1}H+\zeta_{2}H^{2}+\zeta_{3}H^{3}+\zeta_{4}H^{4} .\]

In previous cases the Casimir operator has the form $K=C^{2}+h(A,B)$, where $h(A,B)$ is a polynomial of $A$ and $B$ in a such way that it has the same order as $C^{2}$ in term of momentum.
We thus consider Casimir of the form :
\begin{equation}
K = C^{2}+2c_{1}A^{3}B +2c_{2}A^{2}B+2c_{3}AB^{2}+2c_{4}AB  \label{casimirclas1}
\end{equation} 
\[+c_{5}B^{2}+c_{6}B+c_{7}A^{5}+c_{8}A^{4}+c_{9}A^{3}+c_{10}A^{2}+c_{11}A .\]

Using $\{K,A\}=\{K,B\}=0$ we obtain a set of linear equations in coefficients $c_{i}$ that allows to obtain the solution for the parameter $c_{1}...c_{11}$ in term of the structure constants.

We have the following parameters
\begin{equation}
c_{1}= -\tau,\quad c_{2} =  -\alpha ,\quad c_{3}= -\beta ,\quad c_{4}= -\gamma , \quad c_{5}= -\delta ,\quad c_{6}=   -2 \epsilon, \label{casimirclas2}
\end{equation}
\[ c_{7}=   \frac{2 \lambda }{5} , \quad c_{8}= \frac{\mu }{2} ,\quad c_{9}= \frac{2 \nu }{3},\quad c_{10}= \xi ,\quad c_{11}= 2 \zeta  .\]

This Casimir operators can also to be written in term of the Hamiltonian as a polynomial

\begin{equation}
K= k_{0}+k_{1}H+k_{2}H^{2}+k_{3}H^{3}+k_{4}H^{4}+k_{5}H^{5}.
\end{equation}

\section{Quartic algebras}

In the quantum case, we replace the Poisson bracket by the commutator and the momentum takes the form

\begin{equation}
p_{j}=-i\hbar \frac{\partial}{\partial x_{j}}.
\end{equation}
The integrals $A$ and $B$ are now well defined algebraically and independent quantum mechanical operators. We thus have

\begin{equation}
[H,A]=[H,B]=0.
\end{equation}

The most general quartic quantum algebra is thus :

\begin{subequations}
\begin{equation}
[A,B]=C, \label{quart1}
\end{equation}
\begin{equation}
[A,C]= \tau A^{3} + \alpha A^{2} + \beta \{A,B\} + \gamma A + \delta B + \epsilon , \label{quart2}
\end{equation}
\begin{equation}
[B,C]= \lambda A^{4} + \mu A^{3} + \nu A^{2} + \xi A + \rho B^{2} +\eta B + \omega \{A^{2},B\} + \sigma \{A,B\} +\zeta. \label{quart3}
\end{equation}
\end{subequations}

The Jacobi identity provide also constraint on the structure constants:

\begin{equation}
[A,[B,C]]=[B,[A,C]].  \label{jacobiquant}
\end{equation}

We obtain the relations between the parameters :

\begin{equation}
\omega=-\frac{3}{2} \tau, \quad \sigma=\frac{\beta \tau}{2}-\alpha,\quad \rho=-\beta,\quad \eta=-\gamma+\frac{\delta \tau}{2}.  \label{jacobiquant2}
\end{equation} 

Unlike the cases of quadratic associative and cubic associative algebras the relation obtained differ from the classical case. The quartic algebra takes the form :

\begin{subequations}
\begin{equation}
[A,B]=C, \label{quart1v2}
\end{equation}
\begin{equation}
[A,C]= \tau A^{3} + \alpha A^{2} + \beta \{A,B\} + \gamma A + \delta B + \epsilon , \label{quart2v2}
\end{equation}
\begin{equation}
 [B,C]= \lambda A^{4} + \mu A^{3} + \nu A^{2} + \xi A -\beta B^{2} +(-\gamma+\frac{\delta \tau}{2})B-\frac{3}{2} \tau \{A^{2},B\} + (\frac{\beta \tau}{2}-\alpha) \{A,B\} +\zeta. \label{quart3v2}
\end{equation}
\end{subequations}

In the case of quadratic and cubic algebras the form only differ by a symmetrization/antisymmetrization. Here we see that correction involving $\tau$ appear in lower term. This was observed for the quadratic and cubic case concerning the Casimir operator however for the quartic case correction term appear at the level of the algebra. We recover results for the cubic case by
taking $\tau \rightarrow 0$ and $\lambda \rightarrow 0$. We obtain the quadratic case taking furthermore $ \mu \rightarrow 0$ and the quadratic Racah algebra $QR(3)$ by taking also $\nu \rightarrow 0$.

The Casimir operator has the following form:

\begin{equation}
 K = C^{2}+c_{1}\{A^{3},B\} +c_{2}\{A^{2},B\}+c_{3}\{A,B^{2}\}+c_{4}\{A,B\} \label{casimirquant1}
\end{equation} 
\[+c_{5}B^{2}+c_{6}B+c_{7}A^{5}+c_{8}A^{4}+c_{9}A^{3}+c_{10}A^{2}+c_{11}A . \]

Using $[K,A]=[K,B]=0$, identities given in Appendix A and taking into account ordering of the operators, we obtain a set of equations that allow to obtain the solution for the parameter $c_{1}...c_{11}$ in term of the structure constants.

We have the following parameters
\begin{equation}
c_{1}= -\tau,\quad c_{2} =  -\alpha +\frac{3 \beta  \tau }{2},\quad c_{3}= -\beta ,\quad c_{4}= -\gamma +\beta(\alpha -\frac{\beta  \tau }{2}) , \label{casimirquant2}
\end{equation}
\[c_{5}=  \beta^{2}-\delta ,\quad c_{6}=   -2 \epsilon  +\beta  \gamma  - (1/2) \beta\delta  \tau ,\quad c_{7}=   \frac{2 \lambda }{5} , \quad c_{8}=     -\beta  \lambda +\frac{\mu }{2}  -\frac{9}{4}\tau^{2}  ,  \]
\[c_{9}= (\frac{8 \beta^{2}}{15}+\frac{2 \delta }{3}) \lambda +\frac{2\beta \mu }{3}+\frac{2 \nu }{3}+3\alpha \tau -\frac{3}{2} \beta \tau^{2}, \]
\[c_{10}=   \alpha^{2}+(-\frac{2 \beta^{3}}{15}+\frac{\beta \delta }{3}) \lambda -\frac{\beta^{2} \mu }{6}+\frac{\delta \mu }{2}+\frac{\beta \nu }{3}+\xi +(-\alpha \beta +\frac{3\gamma }{2}) \tau +\frac{1}{4} (\beta^{2} -3 \delta ) \tau^{2}   ,     \]
\[c_{11}=  \alpha \gamma +2 \zeta +(-\frac{2 \beta^{2} \delta }{15}-\frac{\delta^{2}}{15}) \lambda -\frac{\beta \delta  \mu }{6}+\frac{\delta \nu }{3}+(-\frac{\beta \gamma }{2}-\frac{\alpha \delta }{2}) \tau +\frac{1}{4} \beta  \delta \tau^{2} . \]

This Casimir operators can also to be written in term of the Hamiltonian as a polynomial

\begin{equation}
K= k_{0}+k_{1}H+k_{2}H^{2}+k_{3}H^{3}+k_{4}H^{4}+k_{5}H^{5}.
\end{equation}

\subsection{Realizations as deformed oscillator algebras}

In order to obtain the finite-dimensional unitary representations, we consider realizations of the quartic algebras in terms of deformed oscillators algebras \cite{Das8} $\{1,N,b^{\dagger},b\}$ satisfying the following equations :
\begin{equation}
[N,b^{\dagger}]=b^{\dagger},\quad [N,b]=-b,\quad bb^{\dagger}=\Phi(N+1),\quad b^{\dagger}b=\Phi(N),\label{deforosclalg}
\end{equation}
where $\Phi(x)$ is a real function called the structure function satisfying $\Phi(x)=0$ and $\Phi(x)>0$ for $x>0$. We have the existence of Fock type representations when we impose the existence of an integer $p$ such that $\Phi(p+1)=0$. In this case the deformed oscillator algebra is a parafermionic algebra.

We impose the realization of the quartic algebra the form 

\begin{equation}
A=A(N), B=b(N)+b^{\dagger}\rho(N)+\rho(N)b . \label{realiz}
\end{equation}

We found from the first relation~\eqref{quart1v2} of the quartic algebra

\begin{equation}
[A,B]=b^{\dagger}\Delta A(N) \rho(N)-\rho(N) \Delta A(N)b \equiv C . \label{realiz1}
\end{equation}

The second equations of the quartic algebra~\eqref{quart2v2} give us two difference equations to be satisfied by the function $A(N)$ and $b(N)$ :
\begin{equation}
(\Delta A(N))^{2}=\beta (A(N+1)+A(N))+\delta,  \label{realiz2a}
\end{equation}
\begin{equation}
\tau A^{3}(N)+\alpha A^{2}(N)+2 \beta A(N) b(N) +\gamma A(N) +\delta b(N) +\epsilon . \label{realiz2b}
\end{equation}

The following are obtained ( $u$ is a constant determined from the constraints on the structure function ) 

Case 1: $\beta=0$, $\delta \neq 0$
\begin{equation}
A(N)=\sqrt{\delta}(N+u) , \label{realiztype1A}
\end{equation}
\begin{equation}
B(N)=-\sqrt{\delta}\tau (N+u)^{3}-\alpha (N+u)^{2}-\frac{\gamma}{\sqrt{\delta}}(N+u)-\frac{\epsilon}{\delta} . \label{realiztype1B}
\end{equation}

Case 2: $\beta \neq 0$
\begin{equation}
A(N)= (\frac{\beta}{2})(( (N+u)^{2} -(\frac{1}{4})) -( \frac{\delta}{\beta {2}})), \label{realiztype2A}
\end{equation}
\begin{equation}
b(N)=(-\frac{\beta \tau }{8})((N+u)^{2}-(\frac{1}{4}))^{2}+ (\frac{-2\alpha \beta +3 \delta \tau}{8 \beta })((N+u)^{2} - (\frac{1}{4})) \label{realiztype2B}
\end{equation}
\[+( \frac{-4 \beta^{2} \gamma +4 \alpha \beta \delta -3 \delta^{2} \tau }{8 \beta^{3}}) -(\frac{-4 \beta^{2} \gamma \delta +2 \alpha \beta \delta^{2}+8 \beta^{3}\epsilon -\delta^{3} \tau }{8 \beta^{5}})\]
\[( \frac{1}{((N+u)^{2} - (\frac{1}{2}))}).\]

From the third equation~\eqref{quart3v2} of the quartic algebra we have 3 difference equations to be satisfied

\begin{equation}
\Delta A(N)- \Delta A(N+1)=-\beta, \label{realiz3a}
\end{equation}

\begin{equation}
\Delta A(N)(b(N+1)-b(N))=-\beta ( b(N+1)+b(N))+(-\gamma+\frac{\delta \tau}{2})+(A(N+1)+A(N))(-\alpha + \frac{\tau \beta}{2}), \label{realiz3b}
\end{equation}
\[-\frac{3}{2}\tau (A^{2}(N+1)-A^{2}(N))\]

\begin{equation}
- 2 \Phi(N)\rho^{2}(N-1)\Delta A(N-1)+2 \Phi(N+1) \rho^{2}(N)\Delta A(N)=  \label{realiz3c}
\end{equation}
\[-\beta(\Phi(N)\rho^{2}(N-1)+\Phi(N+1)\rho^{2}(N)) -\beta b^{2}(N) +\lambda A^{4}(N)+\mu A^{3}(N) \]
\[+\nu A^{2}(N) +\xi A(N)+\zeta -3 \tau A^{2}(N)b(N)+(-2 \alpha +\tau \beta )A(N)b(N)+(-\gamma +\frac{\delta}{\tau}{2})b(N).  \]

The first two equations i.e.~\eqref{realiz3a} and~\eqref{realiz3b} are satisfied for the two solutions we obtained for $A(N)$ and $b(N)$. This point out how the existence of realizations as deformed oscillators
is connected with the fact that generators satisfied the Jacobi identity.

Using the realizations, the Casimir operator~\eqref{casimirquant1} given by also provide three equations   

\begin{equation}
-\beta (A(N+2)+A(N))+(\beta^{2}-\delta)+\Delta A(N+1) \Delta A(N)=0 ,\label{realiz4a}
\end{equation}

\begin{equation}
-\tau ( A^{3}(N)+A^{3}(N+1))+(-\alpha +\frac{3}{2}\beta \tau )(A^{2}(N)+A^{2}(N+1))-\beta (A(N)b(N)+b(N) A(N+1) \label{realiz4b}
\end{equation}
\[+A(N)b(N+1)+b(N+1)A(N+1))+(-\gamma+\beta \alpha -\frac{\beta^{2}\tau }{2})(A(N)+A(N+1))\]
\[+(\beta^{2}-\delta)(b(N)+b(N+1))+(-2\epsilon +\beta \gamma -\frac{\beta \tau \delta}{2})=0,\]

\begin{equation}
K= - \Phi(N) \Delta A^{2}(N-1) \rho^{2}(N-1) - \Phi(N+1) \Delta A^{2}(N) \rho^{2}(N) \label{realiz4c}
\end{equation}
\[ 2c_{1} A^{3}(N)b(N) +2 c_{2} A^{2}(N)b(N)+ 2 c_{3} A(N) (\Phi \rho^{2}(N-1) +\Phi(N+1)\rho^{2}(N))+2c_{3}A(N)b^{2}(N)\]
\[ 2 c_{4} A(N)b(N) + c_{5}( (\Phi(N) \rho^{2}(N-1) +\Phi(N+1)\rho^{2}(N))+b^{2}(N)) +c_{6}b(N) \]
\[ +c_{7}A^{5}(N) +c_{8}A^{4}(N) +c_{9}A^{3}(N) +c_{10} A^{2}(N) +c_{11}A(N) .\] 

The first two equations i.e.~\eqref{realiz4a} and~\eqref{realiz4b} are satisfied for the two solutions we obtained for $A(N)$ and $b(N)$.

Using the remaining equations~\eqref{realiz3c} and~\eqref{realiz4c} from the third relation of the quartic algebra and the Casimir operator we can found the structure function in term of the arbitrary function $\rho(N)$ that can be chosen in a way that the structure function is a polynomial.

Case 1 :
\begin{equation}
\rho(N)=1,  \label{rhotype1}
\end{equation}

\begin{equation}
\label{phitype1}
\end{equation}
\[ \Phi(N)=-\frac{K}{2 \delta }-\frac{\gamma  \epsilon }{2 \delta^{3/2}}+\frac{\epsilon ^2}{2 \delta^2}-\frac{\zeta }{2 \sqrt{\delta }}+\frac{\epsilon  \tau }{4 \sqrt{\delta }}+\frac{1}{2} N^6 \delta  \tau ^2\]
\[+N (-\frac{\gamma ^2}{2 \delta }+\frac{\alpha  \gamma }{2 \sqrt{\delta }}+\frac{\gamma  \epsilon }{\delta ^{3/2}}-\frac{\alpha  \epsilon }{\delta }+\frac{\zeta }{\sqrt{\delta }}-\frac{1}{30} \delta ^{3/2} \lambda +\frac{\sqrt{\delta } \nu }{6}-\frac{\xi }{2}+\frac{\gamma  \tau }{4}-\frac{1}{4} \alpha  \sqrt{\delta } \tau)\]
\[+N^5 \left(\frac{1}{5} \delta ^{3/2} \lambda +\alpha  \sqrt{\delta } \tau -\frac{3 \delta  \tau^2}{2}\right)+N^4 (\frac{\alpha^2}{2}-\frac{1}{2} \delta^{3/2} \lambda +\frac{\delta  \mu }{4}+\gamma  \tau -\frac{5}{2} \alpha  \sqrt{\delta } \tau -\frac{9 \delta  \tau ^2}{8})\]
\[+N^2 (\frac{\alpha ^2}{2}+\frac{\gamma ^2}{2 \delta }-\frac{3 \alpha  \gamma }{2 \sqrt{\delta }}+\frac{\alpha  \epsilon }{\delta }+\frac{\delta  \mu }{4}-\frac{\sqrt{\delta } \nu }{2}+\frac{\xi }{2}+\frac{3 \gamma  \tau }{4}+\frac{1}{4} \alpha  \sqrt{\delta } \tau -\frac{3 \epsilon  \tau }{2 \sqrt{\delta }}-\frac{3 \delta  \tau ^2}{8})\]
\[+N^3 (-\alpha ^2+\frac{\alpha  \gamma }{\sqrt{\delta }}+\frac{1}{3} \delta ^{3/2} \lambda -\frac{\delta  \mu }{2}+\frac{\sqrt{\delta } \nu }{3}-2 \gamma  \tau +\frac{3}{2} \alpha  \sqrt{\delta } \tau +\frac{\epsilon  \tau }{\sqrt{\delta }}+\frac{\delta  \tau ^2}{4}).\]

Case 2 :
\begin{equation}
\label{rhotype2}
\end{equation}
\[ \rho(N)^{2}= \frac{1}{\sqrt{3932160 (N+u-1)(N+u)\beta^{10}(1-2(N+u))^{2} }}\]

The structure function is a polynomial of order 12 we present in the Appendix B.

These formula coincide with the appropriate limit with those obtained for quadratic and cubic algebra. The Casimir operator $K$ can be written in terms of the Hamiltonian only. We have a energy dependent Fock space of dimension p+1 if
\begin{equation}
\Phi(p+1,u,E)=0, \quad \Phi(0,u,E)=0,\quad \Phi(x)>0, \quad \forall \; x>0 \quad .\label{constraints}
\end{equation}

The Fock space is defined by
\begin{equation*}
H|E,n>=E|E,n>,\quad N|E,n>=n|E,n> \quad b|E,0>=0, \label{fock1}
\end{equation*}
\begin{equation*}
b^{\dagger}|n>=\sqrt{\Phi(n+1,E)}|E,n+1>,\quad b|n>=\sqrt{\Phi(n,E)}|E,n-1>. \label{fock2}
\end{equation*}
The energy $E$ and the constant $u$ are solutions of the equations obtained by the system given by~\eqref{constraints}. They represent the finite-dimensional unitary representations with dimension $p+1$. Let mention that in some cases of systems allowing a cubic algebra, the finite-dimensional unitary representations do not allow to recover all the energy spectrum and the degeneracies \cite{Que4} and one need to consider higher order polynomial algebra \cite{Que5}. In such case, a union of finite-dimensional unitary representations need to be considered for a given energy level. Such phenomena would also be observed for quartic associative algebras.

\section{Example}

In order to illustrate the application of these formula in context of superintegrable systems let us consider an example. There is no classification of superintegrable systems with a second and third order integrals of motion even for the class of systems with separation of variables in Cartesian coordinates on 2D real Euclidean space. Let us apply the results of the previous section on the following Hamiltonian obtained recently \cite{Que4}. The Hamiltonian $H$ is given by

\begin{equation}
H_{x}=-\frac{d^{2}}{dx^{2}}+\frac{x^{2}}{4}+\frac{l(l+1)}{x^{2}}+{4}{1+2l+x^{2}}-\frac{8(1+2l)}{(1+2l +x^{2})^{2}}-1, \label{hamilx}
\end{equation}

\begin{equation}
H_{y}=-\frac{d^{2}}{dy^{2}}+\frac{y^{2}}{4}, \label{hamily}
\end{equation}

\begin{equation}
H=H_{x}+H_{y}, \label{intH}
\end{equation}

the second order integral $A$ takes the form

\begin{equation}
A=H_{x}-H_{y}, \label{intA}
\end{equation}

and the fourth order integral $B$ is

\begin{equation}
B=\frac{1}{16(x+2 l x+x^{3})^{4}}(-256 l^8-256 l^{7}(4+3 x^{2})-128 l^{6}(1+21 x^{2}+7 x^{4})-64 l^{5}(-50-6 x^{2}+42 x^{4}+7 x^{6}) \label{intB}
\end{equation}
\[-80 l^{4}(-59-96 x^{2}-10 x^{4}+14 x^{6})+16 l^{3}(182+519 x^{2}+380 x^{4}+36 x^{6}+7 x^{10})\]
\[+x^{4} (192-1662 x^{2}+201 x^{4}+16 x^{6}+14 x^{8}+6 x^{10}+x^{12})+8 l^{2}(106+429 x^{2}+648 x^{4}\]
\[+248 x^{6}+33 x^{8}+21 x^{10}+7 x^{12})+4 k (24+126 x^{2}+424 x^{4}-647 x^{6}+66 x^{8}+22 x^{10}+14 x^{12}+3 x^{14})) y \]
\[ +\frac{1}{8 x^{2} (1+2 l+x^{2})^{3}}((32 l^{5}+16 l^{4} (5+x^{2})-8 l^{3}(-9-4 x^{2}+6 x^{4})-4 l^{2}(-7-35 x^{2}+18 x^{4}+14 x^{6})\]
\[+x^{2}(30-153 x^{2}+x^{4}-11 x^{6}-3 x^{8})-2 l(-2-62 x^{2}+165 x^{4}+28 x^{6}+11 x^{8})) \frac{\partial}{\partial y} )-\frac{4 l y}{x^{3}}\frac{\partial}{\partial x}\]
\[-\frac{4 l^{2} y}{x^{3}}\frac{\partial}{\partial x}-\frac{64 x^{3} y}{(1+2 l+x^{2})^{3}}\frac{\partial}{\partial x}+\frac{44 x y }{(1+2 l+x^{2})^{2}}\frac{\partial}{\partial x}-\frac{8 l x y}{(1+2 l+x^{2})^{2}}\frac{\partial}{\partial x}\]
\[-\frac{4 x^{3} y }{(1+2 l+x^{2})^{2}}\frac{\partial}{\partial x}+\frac{4 x y}{1+2 l+x^{2}}\frac{\partial}{\partial x}-\frac{l}{x}\frac{\partial}{\partial y}\frac{\partial}{\partial x}-\frac{l^{2}}{x}\frac{\partial}{\partial y}\frac{\partial}{\partial x}\]
\[+\frac{3}{2} x \frac{\partial}{\partial y}\frac{\partial}{\partial x}-l x \frac{\partial}{\partial y}\frac{\partial}{\partial x}-\frac{1}{4} x^{3} \frac{\partial}{\partial y}\frac{\partial}{\partial x}-\frac{12 x^{3}}{(1+2 l+x^{2})^{2}}\frac{\partial}{\partial y}\frac{\partial}{\partial x}\]
\[+\frac{6 x}{1+2 l+x^{2}}\frac{\partial}{\partial y}\frac{\partial}{\partial x}-\frac{4 l x}{1+2 l+x^{2}}\frac{\partial}{\partial y}\frac{\partial}{\partial x}-\frac{2 x^{3}}{1+2 l+x^{2}}\frac{\partial}{\partial y}\frac{\partial}{\partial x}-\frac{3}{2} y \frac{\partial^{2}}{\partial x^{2}}+l y \frac{\partial^{2}}{\partial x^{2}}+\frac{2 l y}{x^2}\frac{\partial^{2}}{\partial x^{2}}+\frac{2 l^{2} y}{x^{2}}\frac{\partial^{2}}{\partial x^{2}}\]
\[+\frac{16 x^{2} y}{(1+2l+x^{2})^{2}}\frac{\partial^{2}}{\partial x^{2}} -\frac{6 y}{1+2 l+x^{2}}\frac{\partial^{2}}{\partial x^{2}}+\frac{4 l y}{1+2 l+x^{2}}\frac{\partial^{2}}{\partial x^{2}}+\frac{2 x^{2} y}{1+2 l+x^{2}}\frac{\partial^{2}}{\partial x^{2}}\]
\[+\frac{3}{2} \frac{\partial}{\partial y}\frac{\partial^{2}}{\partial x^{2}}+x \frac{\partial}{\partial y}\frac{\partial^{3}}{\partial x^{3}}-y \frac{\partial^{4}}{\partial x^{4}}.\]

The structure constants of the quartic algebra are given by

\begin{equation}
\delta=16,\quad \lambda=-\frac{5}{2},\quad \mu=-3H-(10+4l), \label{param}
\end{equation}
\[ \nu=-\frac{3}{2}H^{2}-(15+6l)H-(25+18l),\]
\[ \xi=H^{3}-(22 +12 l)H-(35+34l-12l^{2}-8l^{3}), \]
\[ \zeta= \frac{3}{4}H^{4}+(5+2l)H^{3}+(3+6l)H^{2}-(20+8l)H-(\frac{1}{4}(175+160-8l^{2}-64l^{3}16l^{4}).\]

We can calculate the Casimir operator and write this operator only in term of the Hamiltonian

\begin{equation}
K=-\frac{1}{2}H^{5}-(5+2l)H^{4}-(14+12l)H^{3}+(-1+2l)^{2}(5+2l)H^{2} \label{casimirhamil}
\end{equation}
\[+\frac{1}{2}(149 +32l +8l^{2}+64l^{3}+16l^{4})H-(4-3+2l)(1+2l)(5+2l).\]

We can calculate the structure function

\begin{equation}
\Phi(H,u,x)=\frac{1}{64}(2+H-4(x+u))(-1+H-2l+4(x+u)) \label{strufunc1}
\end{equation}
\[(-3+H+2l+4(x+u))(1+H+2l+4(x+u))(5+H+2l+4(x+u)).\]

Thus

\begin{equation}
\Phi(E,u,x)=-16(x+u-(\frac{2+E}{4}))(x+u-(\frac{1}{4}(-5+E-2l))) \label{strufunc2}
\end{equation}
\[(x+u-(\frac{1}{4}(-1-E-2l)))(x+u-(\frac{1}{4}(3-E-2l)))(x+u-(\frac{1}{4}(1-E+2l))).\]

From $\Phi(E,u,0)=0$ we obtain

\begin{equation}
u_{1}=\frac{2+E}{4},\quad u_{2}=\frac{1}{4}(-5-E-2l), \quad u_{3}=\frac{1}{4}(-1-E-2l), \label{paramu1}
\end{equation}

\begin{equation}
u_{4}=\frac{1}{4}(3-E-2l), \quad u_{5}=\frac{1}{4}(1-E+2l) .
\end{equation}

Using $u_{5}$ the structure function takes the form

\begin{equation}
\Phi(x)=\frac{1}{2}(1+2E-2l-4x)x(-1+2l+2x)(1+2l+2x)(3+2l+2x). \label{strufunc3}
\end{equation}

Using $\Phi(E,u_{5},p+1)=0$ we found the degenerate energy spectrum

\begin{equation}
E=2p+l+\frac{3}{2},  \label{strufuncenerg}
\end{equation}

with

\begin{equation}
\Phi=2x(p+1-x)(2x+2l-1)(2x+2l+1)(2x+2l+3). \label{strufunc4}
\end{equation}
 
This result corroborate the one obtained using ladder operators \cite{Que4}. 

\section{Conclusion}

We obtained the most general quartic Poisson and quartic associative algebras respectively for classical and quantum superintegrable systems with a second and a fourth order integrals of motion. In quantum mechanics, this is a deformation of the quadratic Racah algebra $QR(3)$. Unlike for the case of the quadratic and cubic algebras, in the case of quartic algebras the classical and quantum cases differ when we impose the Jacobi identity.

We present the realizations in terms of deformed oscillator algebras of the quantum quartic algebra. We discuss how these results can be used to obtain finite dimensional unitary representations. This allow also to provide a method to obtain the energy spectrum of superintegrable with a second and a fourth order integrals of motion. The classification of superintegrable systems beyond quadratically superintegrable systems is difficult and only specific examples with second and fourth order integrals are known and a classification need to be started for such systems. 

Recently, a classification of superintegrable systems and the representation theory of their quadratic algebra was related to various orthogonal polynomials as Racah, Wilson, Hahn, Jacobi, Bessel, Krawtchouk, Meixner-Pollaczek, Gegenbauer, Laguerre, Hermite, Tchebicheff and Charlier. Moreover, specific examples of superintegrable systems with higher integrals of motion were related with exceptional orthogonal polynomial \cite{Pos3}. These results point out how superintegrable systems and their algebraic structures are closely related with orthogonal polynomials and thus the study of the most general quartic associative algebra could provide new connections with orthogonal polnyomials.

In addition, in recent years various results for quadratic algebras in three dimensions \cite{Mil2,Mil3,Tan1,Das9,Das10,Mar6} and five dimensions with a non Abelian monopole were obtained \cite{Mar7} and thus the study of polynomial algebras and their realizations could be extended in three dimensions.

These algebraic structures are constructed in context of superintegrability, however they are objects interesting by themselves and as Lie algebra they can found applications in other contexts in physics or mathematical physics. Recently some classes of polynomial algebras found applications in condensed matter \cite{Zha1,Zha2,Lin1}. 

The quadratic Racah algebra $QR(3)$ algebra can be extended to the quadratic Askey-Wilson algebra denoted by $QAW(3)$ \cite{Gra1} by replacing the commutator by the deformed commutator  $[A,B]_{\omega}=e^{\omega}A-B e^{-\omega}$. 
A generalization of the Askey-Wilson algebra with cubic term and (p,q)-deformation was studied GAW(3) \cite{Lav1}. The investigation of q-deformation of these polynomial associative algebras could also be studied.

\textbf{Acknowledgments} 

The research of I.M. was supported by the Australian Research Council through Discovery Project DP110101414. He thanks Phil Isaac for very interesting discussions.

\section{Appendix A}

Let us present the following list of indentities that allow to solve the constraint from the Jacobi identity and calculate the Casimir operator.

\begin{flushleft}
1. $[A,B]=C$ \\
2. $[ A^{2},B]=\{C,A\}$ \\
3. $[A^{3},B]= \{C,A^{2}\}+\frac{1}{2}(5)$\\
4. $[A^{4},B]=\{C,A^{3}\}+(6)$\\
5. $2ACA= \{C,A^{2}\}-\beta \{C,A\}-\delta C $\\
6. $ A^{2}CA+ACA^{2}=\{C,A^{3}\}-2\beta \{C,A^{2}\}+(\beta^{2}-\delta)\{A,C\}+\beta \delta C$\\
7. $[B,\{A,B\}]=-\{C,B\}$\\
8. $[\{A,B\},A]=\{C,A\}$\\
9. $[\{A,B\},A^{2}]= - \{C,A^{2}\}-(5)$\\
10. $[\{A,B\},A^{3}]=\frac{3}{2}\{C,A^{3}\}+\frac{3}{2}(6)-\frac{\beta}{2}\{C,A^{2}\}-\frac{\beta}{2}(5)-\frac{\delta}{2}\{C,A\}$\\
11. $A^{3}CA+ACA^{3}=\{C,A^{4}\}-\delta (3)+\beta (10) $\\
12. $[A^{5},B]=\{C,A^{4}\}+(11)+\frac{1}{2}(13)$
13. $2A^{2}CA^{2}=(11)+\beta (6) -\delta (5)$\\
14. $[\{A^{2},B\},A]=-\{C,A^{2}\}$\\
15. $[\{A^{3},B\},A]=-\{C,A^{3}\}$\\
16. $[B^{2},A]=\{C,B\}$\\
17. $\{A,\{B,C\}\}=\{C,\{A,B\}\}+\tau (3) +\alpha (2) -\beta (7) +\gamma C$\\
18. $[\{A^{3},B\},B]=\frac{3}{2}(20)-\frac{\beta}{2}(24) -\frac{\delta}{2}\{C,B\}$\\
19. $[\{A^{2},B\},B]=(24)$\\
20. $\{B,\{C,A^{2}\}\}=\{C,\{A^{2},B\}\}+\beta (21)- (-\gamma +\frac{\delta \tau}{2})(2)-\frac{3\tau}{2}(23) -(-\alpha+\frac{\tau \beta}{2}) (22)$\\
21. $[A^{2},B^{2}]=\{C,\{A,B\}\}+\tau (3)+\alpha (2) -\beta (7) + \gamma (1)$\\
22. $[A^{2},\{A,B\}]=\{C,A^{2}\}+(5) $\\
23. $[A^{2},\{A^{2},B\}]=\{C,A^{3}\}+(6)$\\
24. $\{B,\{C,A\}\}=\{C,\{A,B\}\}+\beta \{C,B\}-(-\gamma+\frac{\delta \tau}{2})(1)-\frac{3\tau }{2}(14)+(-\alpha +\frac{\tau \beta}{2})(8)$\\
25. $[C^{2},A]=\tau\{C,A^{3}\}-\alpha\{C,A^{2}\}-\beta \{C,\{A,B\}\}-\gamma\{C,A\}-\delta\{C,B\}-2\epsilon C$\\
26. $[\{A,B^{2}\},A]=-\{A,\{C,B\}\}=-(17)$\\
27. $[C^{2},B]=-\lambda\{C,A^{4}\}-\mu\{C,A^{3}\}-\nu\{C,A^{2}\}-\xi \{C,A\}+\beta \{C,B^{2}\}-(-\gamma+\frac{\delta \tau}{2})\{C,B\}$\\
$+\frac{3}{2}\tau \{C,\{A^{2},B\}\}-(-\alpha +\frac{\tau \beta}{2})\{C,\{A,B\}\}-2\zeta C$\\
28. $[\{A,B^{2}\},B]=\{C,B^{2}\}$\\
29. $[\{A,B\},B]=\{C,B\}$\\
\end{flushleft}

Let us denote that many of the commutators can be written only in term of the anticommutator of the form $\{C,A^{i}\}$. We have also the following relations

\[ [K,A]=(25)+c_{1}(15)+c_{2}(14)=c_{3}(26)+c_{4}(8)+c_{5}(16)+c_{6}(-(1))\]
\[ [K,B]=(27)+c_{1}(18)+c_{2}(19)+c_{3}(28)+c_{4}(21)+c_{7}(12)+c_{8}(4)+c_{9}(3)+c_{10}(2)+c_{11}(1)\]

\section{Appendix B}

Let us present the structure function for the second type of realization of the quartic algebra.

\[\Phi(N)=-983040 K \beta ^8+8640 \alpha ^2 \beta ^{10}-46080 \alpha  \beta ^9 \gamma -184320 \beta ^8 \gamma ^2+46080 \alpha ^2 \beta ^8 \delta +61440 \alpha  \beta ^7 \gamma  \delta\]
\[ -491520 \beta ^6 \gamma ^2 \delta -30720 \alpha ^2 \beta ^6 \delta ^2+737280 \alpha  \beta ^5 \gamma  \delta ^2+983040 \beta ^4 \gamma ^2 \delta ^2-245760 \alpha ^2 \beta ^4 \delta ^3\]
\[-983040 \alpha  \beta ^3 \gamma  \delta ^3+245760 \alpha ^2 \beta ^2 \delta ^4-368640 \alpha  \beta ^8 \epsilon +983040 \beta ^7 \gamma  \epsilon -983040 \alpha  \beta ^6 \delta  \epsilon\]
\[ -3932160 \beta ^5 \gamma  \delta  \epsilon +1966080 \alpha  \beta ^4 \delta ^2 \epsilon +3932160 \beta ^6 \epsilon ^2+245760 \beta ^9 \zeta -983040 \beta ^7 \delta  \zeta\]
\[ -3204 \beta ^{13} \lambda -10608 \beta ^{11} \delta  \lambda +3968 \beta ^9 \delta ^2 \lambda -50688 \beta ^7 \delta ^3 \lambda -128000 \beta ^5 \delta ^4 \lambda\]
\[ -12288 \beta ^3 \delta ^5 \lambda -4680 \beta ^{12} \mu -17280 \beta ^{10} \delta  \mu +6400 \beta ^8 \delta ^2 \mu +10240 \beta ^6 \delta ^3 \mu +30720 \beta ^4 \delta ^4 \mu \]
\[+11520 \beta ^{11} \nu +46080 \beta ^9 \delta  \nu -20480 \beta ^7 \delta ^2 \nu -81920 \beta ^5 \delta ^3 \nu -46080 \beta ^{10} \xi -122880 \beta ^8 \delta  \xi\]
\[ +245760 \beta ^6 \delta ^2 \xi -15120 \alpha  \beta ^{11} \tau +40320 \beta ^{10} \gamma  \tau -66240 \alpha  \beta ^9 \delta  \tau +153600 \beta ^8 \gamma  \delta  \tau\]
\[ -23040 \alpha  \beta ^7 \delta ^2 \tau -184320 \beta ^6 \gamma  \delta ^2 \tau +92160 \alpha  \beta ^5 \delta ^3 \tau -491520 \beta ^4 \gamma  \delta ^3 \tau +307200 \alpha  \beta ^3 \delta ^4 \tau\]
\[ +491520 \beta ^2 \gamma  \delta ^4 \tau -245760 \alpha  \beta  \delta ^5 \tau +322560 \beta ^9 \epsilon  \tau +552960 \beta ^7 \delta  \epsilon  \tau +737280 \beta ^5 \delta ^2 \epsilon  \tau\]
\[ -983040 \beta ^3 \delta ^3 \epsilon  \tau +5535 \beta ^{12} \tau ^2+5400 \beta ^{10} \delta  \tau ^2-115440 \beta ^8 \delta ^2 \tau ^2-264960 \beta ^6 \delta ^3 \tau ^2\]
\[-311040 \beta ^4 \delta ^4 \tau ^2-92160 \beta ^2 \delta ^5 \tau ^2+61440 \delta ^6 \tau ^2\]
\[+N (3932160 K \beta ^8-46080 \alpha ^2 \beta ^{10}+307200 \alpha  \beta ^9 \gamma +491520 \beta ^8 \gamma ^2-307200 \alpha ^2 \beta ^8 \delta +491520 \alpha  \beta ^7 \gamma  \delta\]
\[ +1966080 \beta ^6 \gamma ^2 \delta -245760 \alpha ^2 \beta ^6 \delta ^2-2949120 \alpha  \beta ^5 \gamma  \delta ^2+983040 \alpha ^2 \beta ^4 \delta ^3+983040 \alpha  \beta ^8 \epsilon\]
\[ -3932160 \beta ^7 \gamma  \epsilon +3932160 \alpha  \beta ^6 \delta  \epsilon -1966080 \beta ^9 \zeta +3932160 \beta ^7 \delta  \zeta +12576 \beta ^{13} \lambda +38592 \beta ^{11} \delta  \lambda \]
\[-38912 \beta ^9 \delta ^2 \lambda +141312 \beta ^7 \delta ^3 \lambda +450560 \beta ^5 \delta ^4 \lambda +49152 \beta ^3 \delta ^5 \lambda +20640 \beta ^{12} \mu +92160 \beta ^{10} \delta  \mu \]
\[+66560 \beta ^8 \delta ^2 \mu +81920 \beta ^6 \delta ^3 \mu -122880 \beta ^4 \delta ^4 \mu -61440 \beta ^{11} \nu -307200 \beta ^9 \delta  \nu -163840 \beta ^7 \delta ^2 \nu\]
\[ +327680 \beta ^5 \delta ^3 \nu +307200 \beta ^{10} \xi +983040 \beta ^8 \delta  \xi -983040 \beta ^6 \delta ^2 \xi +72000 \alpha  \beta ^{11} \tau -245760 \beta ^{10} \gamma  \tau\]
\[ +384000 \alpha  \beta ^9 \delta  \tau -1105920 \beta ^8 \gamma  \delta  \tau +522240 \alpha  \beta ^7 \delta ^2 \tau +245760 \alpha  \beta ^5 \delta ^3 \tau +1966080 \beta ^4 \gamma  \delta ^3 \tau \]
\[-1228800 \alpha  \beta ^3 \delta ^4 \tau -675840 \beta ^9 \epsilon  \tau -1474560 \beta ^7 \delta  \epsilon  \tau -2949120 \beta ^5 \delta ^2 \epsilon  \tau -23400 \beta ^{12} \tau ^2\]
\[-38880 \beta ^{10} \delta  \tau ^2+372480 \beta ^8 \delta ^2 \tau ^2+844800 \beta ^6 \delta ^3 \tau ^2+1013760 \beta ^4 \delta ^4 \tau ^2+368640 \beta ^2 \delta ^5 \tau^2)\]
\[+N^2 (-3932160 K \beta ^8+15360 \alpha ^2 \beta ^{10}-552960 \alpha  \beta ^9 \gamma +491520 \beta ^8 \gamma ^2+552960 \alpha ^2 \beta ^8 \delta -3440640 \alpha  \beta ^7 \gamma  \delta\]
\[ -1966080 \beta ^6 \gamma ^2 \delta +1720320 \alpha ^2 \beta ^6 \delta ^2+2949120 \alpha  \beta ^5 \gamma  \delta ^2-983040 \alpha ^2 \beta ^4 \delta ^3+983040 \alpha  \beta ^8 \epsilon\]
\[ +3932160 \beta ^7 \gamma  \epsilon -3932160 \alpha  \beta ^6 \delta  \epsilon +5898240 \beta ^9 \zeta -3932160 \beta ^7 \delta  \zeta +18976 \beta ^{13} \lambda +72512 \beta ^{11} \delta  \lambda \]
\[+346112 \beta ^9 \delta ^2 \lambda +473088 \beta ^7 \delta ^3 \lambda -204800 \beta ^5 \delta ^4 \lambda -49152 \beta ^3 \delta ^5 \lambda +19040 \beta ^{12} \mu -30720 \beta ^{10} \delta  \mu\]
\[ -250880 \beta ^8 \delta ^2 \mu -573440 \beta ^6 \delta ^3 \mu +122880 \beta ^4 \delta ^4 \mu +20480 \beta ^{11} \nu +552960 \beta ^9 \delta  \nu +1146880 \beta ^7 \delta ^2 \nu\]
\[ -327680 \beta ^5 \delta ^3 \nu -552960 \beta ^{10} \xi -2949120 \beta ^8 \delta  \xi +983040 \beta ^6 \delta ^2 \xi +27840 \alpha  \beta ^{11} \tau +307200 \beta ^{10} \gamma  \tau\]
\[ -353280 \alpha  \beta ^9 \delta  \tau +2580480 \beta ^8 \gamma  \delta  \tau -1628160 \alpha  \beta ^7 \delta ^2 \tau +2949120 \beta ^6 \gamma  \delta ^2 \tau -2703360 \alpha  \beta ^5 \delta ^3 \tau\]
\[ -1966080 \beta ^4 \gamma  \delta ^3 \tau +1228800 \alpha  \beta ^3 \delta ^4 \tau -1536000 \beta ^9 \epsilon  \tau -1474560 \beta ^7 \delta  \epsilon  \tau +2949120 \beta ^5 \delta ^2 \epsilon  \tau \]
\[-21000 \beta ^{12} \tau ^2+96480 \beta ^{10} \delta  \tau ^2+433920 \beta ^8 \delta ^2 \tau ^2+814080 \beta ^6 \delta ^3 \tau ^2-92160 \beta ^4 \delta ^4 \tau ^2-368640 \beta ^2 \delta ^5 \tau ^2)\]
\[+N^3 (307200 \alpha ^2 \beta ^{10}-491520 \alpha  \beta ^9 \gamma -1966080 \beta ^8 \gamma ^2+491520 \alpha ^2 \beta ^8 \delta +5898240 \alpha  \beta ^7 \gamma  \delta \]
\[-2949120 \alpha ^2 \beta ^6 \delta ^2-3932160 \alpha  \beta ^8 \epsilon -7864320 \beta ^9 \zeta -120448 \beta ^{13} \lambda -367616 \beta ^{11} \delta  \lambda -860160 \beta ^9 \delta ^2 \lambda\]
\[ -1720320 \beta ^7 \delta ^3 \lambda -491520 \beta ^5 \delta ^4 \lambda -197120 \beta ^{12} \mu -614400 \beta ^{10} \delta  \mu -368640 \beta ^8 \delta ^2 \mu +983040 \beta ^6 \delta ^3 \mu\]
\[ +409600 \beta ^{11} \nu+491520 \beta ^9 \delta  \nu -1966080 \beta ^7 \delta ^2 \nu -491520 \beta ^{10} \xi +3932160 \beta ^8 \delta  \xi -599040 \alpha  \beta ^{11} \tau \]
\[+860160 \beta ^{10} \gamma  \tau -1781760 \alpha  \beta ^9 \delta  \tau -983040 \beta ^8 \gamma  \delta  \tau -245760 \alpha  \beta ^7 \delta ^2 \tau -5898240 \beta ^6 \gamma  \delta ^2 \tau \]
\[+4915200 \alpha  \beta ^5 \delta ^3 \tau +3440640 \beta ^9 \epsilon  \tau +5898240 \beta ^7 \delta  \epsilon  \tau +204000 \beta ^{12} \tau ^2+69120 \beta ^{10} \delta  \tau ^2\]
\[-1981440 \beta ^8 \delta ^2 \tau ^2-4300800 \beta ^6 \delta ^3 \tau ^2-1843200 \beta ^4 \delta ^4 \tau ^2)\]
\[+N^4 (-522240 \alpha ^2 \beta ^{10}+2703360 \alpha  \beta ^9 \gamma +983040 \beta ^8 \gamma ^2-2703360 \alpha ^2 \beta ^8 \delta -2949120 \alpha  \beta ^7 \gamma  \delta\]
\[ +1474560 \alpha ^2 \beta ^6 \delta ^2+1966080 \alpha  \beta ^8 \epsilon +3932160 \beta ^9 \zeta +12608 \beta ^{13} \lambda -77312 \beta ^{11} \delta  \lambda -307200 \beta ^9 \delta ^2 \lambda\]
\[ +614400 \beta ^7 \delta ^3 \lambda +245760 \beta ^5 \delta ^4 \lambda +136960 \beta ^{12} \mu +1044480 \beta ^{10} \delta  \mu +2027520 \beta ^8 \delta ^2 \mu -491520 \beta ^6 \delta ^3 \mu \]
\[-696320 \beta ^{11} \nu -2703360 \beta ^9 \delta  \nu +983040 \beta ^7 \delta ^2 \nu +2703360 \beta ^{10} \xi -1966080 \beta ^8 \delta  \xi +622080 \alpha  \beta ^{11} \tau \]
\[-2396160 \beta ^{10} \gamma  \tau +3962880 \alpha  \beta ^9 \delta  \tau -4423680 \beta ^8 \gamma  \delta  \tau +6266880 \alpha  \beta ^7 \delta ^2 \tau +2949120 \beta ^6 \gamma  \delta ^2 \tau \]
\[-2457600 \alpha  \beta ^5 \delta ^3 \tau +737280 \beta ^9 \epsilon  \tau -2949120 \beta ^7 \delta  \epsilon  \tau -150000 \beta ^{12} \tau ^2-933120 \beta ^{10} \delta  \tau ^2\]
\[-1313280 \beta ^8 \delta ^2 \tau ^2+1290240 \beta ^6 \delta ^3 \tau ^2+921600 \beta ^4 \delta ^4 \tau ^2)\]
\[+N^5 (-245760 \alpha ^2 \beta ^{10}-2949120 \alpha  \beta ^9 \gamma +2949120 \alpha ^2 \beta ^8 \delta +373760 \beta ^{13} \lambda +1165312 \beta ^{11} \delta  \lambda \]
\[+2457600 \beta ^9 \delta ^2 \lambda +1474560 \beta ^7 \delta ^3 \lambda +547840 \beta ^{12} \mu +491520 \beta ^{10} \delta  \mu -2211840 \beta ^8 \delta ^2 \mu -327680 \beta ^{11} \nu \]
\[+2949120 \beta ^9 \delta  \nu -2949120 \beta ^{10} \xi +1259520 \alpha  \beta ^{11} \tau +983040 \beta ^{10} \gamma  \tau -245760 \alpha  \beta ^9 \delta  \tau +5898240 \beta ^8 \gamma  \delta  \tau \]
\[-7372800 \alpha  \beta ^7 \delta ^2 \tau -2949120 \beta ^9 \epsilon  \tau -441600 \beta ^{12} \tau ^2\]
\[+1428480 \beta ^{10} \delta  \tau ^2+6819840 \beta ^8 \delta ^2 \tau ^2+3686400 \beta ^6 \delta ^3 \tau ^2)\]
\[+N^6(1228800 \alpha ^2 \beta ^{10}+983040 \alpha  \beta ^9 \gamma -983040 \alpha ^2 \beta ^8 \delta -185344 \beta ^{13} \lambda -161792 \beta ^{11} \delta  \lambda \]
\[-491520 \beta ^9 \delta ^2 \lambda -491520 \beta ^7 \delta ^3 \lambda -803840 \beta ^{12} \mu -2457600 \beta ^{10} \delta  \mu +737280 \beta ^8 \delta ^2 \mu +1638400 \beta ^{11} \nu\]
\[ -983040 \beta ^9 \delta  \nu +983040 \beta ^{10} \xi -2426880 \alpha  \beta ^{11} \tau +1966080 \beta ^{10} \gamma  \tau -5652480 \alpha  \beta ^9 \delta  \tau \]
\[-1966080 \beta ^8 \gamma  \delta  \tau +2457600 \alpha  \beta ^7 \delta ^2 \tau +983040 \beta ^9 \epsilon  \tau +764160 \beta ^{12} \tau ^2\]
\[+1428480 \beta ^{10} \delta  \tau ^2-2396160 \beta ^8 \delta ^2 \tau ^2-1228800 \beta ^6 \delta ^3 \tau ^2)\]
\[+N^7 (-983040 \alpha ^2 \beta ^{10}-446464 \beta ^{13} \lambda -1556480 \beta ^{11} \delta  \lambda -1966080 \beta ^9 \delta ^2 \lambda -204800 \beta ^{12} \mu \]
\[+1966080 \beta ^{10} \delta  \mu -1310720 \beta ^{11} \nu +245760 \alpha  \beta ^{11} \tau -1966080 \beta ^{10} \gamma  \tau\]
\[ +4915200 \alpha  \beta ^9 \delta  \tau -291840 \beta ^{12} \tau ^2-4792320 \beta ^{10} \delta  \tau ^2-3686400 \beta ^8 \delta ^2 \tau ^2)\]
\[+N^8 (245760 \alpha ^2 \beta ^{10}+123904 \beta ^{13} \lambda +20480 \beta ^{11} \delta  \lambda +491520 \beta ^9 \delta ^2 \lambda +972800 \beta ^{12} \mu -491520 \beta ^{10} \delta  \mu \]
\[+327680 \beta ^{11} \nu +1781760 \alpha  \beta ^{11} \tau +491520 \beta ^{10} \gamma  \tau\ -1228800 \alpha  \beta ^9 \delta  \tau\]
\[ -618240 \beta ^{12} \tau ^2+1751040 \beta ^{10} \delta  \tau ^2+921600 \beta ^8 \delta ^2 \tau ^2)\]
\[+N^9 (368640 \beta ^{13} \lambda +1228800 \beta ^{11} \delta  \lambda -614400 \beta ^{12} \mu -1228800 \alpha  \beta ^{11} \tau +1259520 \beta ^{12} \tau ^2+1843200 \beta ^{10} \delta  \tau ^2)\]
\[+N^{10} (73728 \beta ^{13} \lambda -245760 \beta ^{11} \delta  \lambda +122880 \beta ^{12} \mu +245760 \alpha  \beta ^{11} \tau -460800 \beta ^{12} \tau ^2-368640 \beta ^{10} \delta  \tau ^2)\]
\[+N^{11} (-294912 \beta ^{13} \lambda -368640 \beta ^{12} \tau ^2)\]
\[+N^{12} (49152 \beta ^{13} \lambda +61440 \beta ^{12} \tau ^2)\]


\end{document}